**Bismuth vanadate layers alternated with nanoparticle-doped silicon dioxide layers for one-dimensional multilayer photonic crystals**

Francesco Scotognella [1,2,*]
[1] Dipartimento di Fisica, Politecnico di Milano, piazza Leonardo da Vinci 32, 20133 Milano, Italy
[2] Center for Nano Science and Technology@PoliMi, Istituto Italiano di Tecnologia (IIT), Via Giovanni Pascoli, 70/3, 20133, Milan, Italy
**\*** Correspondence: francesco.scotognella@polimi.it;

**Abstract:** Bismuth vanadate is one of most studied materials in the field of photocatalysis due to its high photocatalytic activity. The modulation of its optical properties can be engineered by fabricating photonic crystals in which bismuth vanadate is included. In the present study, bismuth vanadate in the monoclinic clinobisvanite structure has been included in a one-dimensional photonic crystal. Bismuth vanadate layers are alternated with layers on silicon dioxide doped with titanium dioxide nanoparticles and silver nanoparticles. The optical properties of the crystal have been studied by employing the transfer matrix method, taking into account all the refractive index dispersions of the selected materials. The transmission spectra have been studied as a function of the bismuth vanadate layers thickness and the filling factor of titanium dioxide and silver dioxide in the silicon dioxide layers.

**Introduction**

Bismuth vanadate ($BiVO_4$) is a very interesting material for photocatalysis [1]. The first pioneering works on the photocatalytic activity of $BiVO_4$ have been reported by Kudo et al. in 1998 [2] and in 1999 [3]. It has been understood that $BiVO_4$ in the monoclinic clinobisvanite structure shows a remarkable photocatalytic activity [2–5]. There are also other interesting structures of $BiVO_4$: In 2014 Cooper et al. have studied the electronic properties of monoclinic sheelite bismuth vanadate [6].

To optimize the employment of $BiVO_4$ in devices it is very important to know its optical properties. Zhao, Li, and Zou have reported a theoretical analysis in which they have determined the complex dielectric function of monoclinic clinobisvanite $BiVO_4$ [7]. In 2014 Sarkar, Das, and Chattopadhyay have experimentally measured the complex refractive index dispersion, between 300 and 800 nm, of a monocrystalline monoclinic thin film of $BiVO_4$ fabricated via radiofrequency sputtering [8].

An interesting way to combine light manipulation and confinement and the optical properties of $BiVO_4$ is the inclusion of $BiVO_4$ in a photonic crystal [9]. Photonic crystals are composite materials in which the alternation of two materials with different refractive indexes has a periodicity that is comparable with the wavelength of light [10–12]. Photonic crystals can show such periodicity in one, two and three dimensions [10,13]. The one-dimensional (1D) case is a very flexible way, in terms of fabrication and choice of materials, to fabricate photonic crystals. Very high quality metal oxide-based 1D photonic crystals have been fabricated via radiofrequency sputtering [14]. Via spin coating it is possible to fabricate 1D photonic crystals with polymers [15–18], sol-gel materials [19], and metal oxide nanoparticles [20–22].

In this work the optical properties of 1D photonic crystals that are composed by layers of monoclinic clinobisvanite $BiVO_4$ and layers of $SiO_2$ doped with $TiO_2$ nanoparticles and Ag nanoparticles have been studied. The transfer matrix method has been employed, taking into account the complex refractive index dispersions of all the materials considered in the structure. The versatility of the 1D crystal has been highlighted by reporting the tunability of the photonic band gap as a function of the thickness of $BiVO_4$ layers, the filling factors of $TiO_2$ and Ag nanoparticles in the $SiO_2$ layers.

**Methods**

Zhao et al. [7] report the complex function of monoclinic clinobisvanite BiVO4. In this work, the direction orthogonal to (010) planes has been selected (b-axis in Ref. [7]). In this work the (010) direction has been selected since for this direction the imaginary part of the dielectric constant is related to a broader transparency (up to 400 nm). The complex refractive index dispersion $n_{BiVO_4}$ has been determined by using

$$n_{real} = \sqrt{(|\sqrt{\varepsilon_1^2 + \varepsilon_2^2}| + \varepsilon_1)/2} \;;\; n_{imag} = \sqrt{(|\sqrt{\varepsilon_1^2 + \varepsilon_2^2}| - \varepsilon_1)/2} \tag{1}$$

being $\varepsilon_1$ the real part of the dielectric function and $\varepsilon_2$ the imaginary part of the dielectric function. For the 3 component-layer, a layer of SiO2 doped with TiO2 nanoparticles and Ag nanoparticles has been considered. The Maxwell-Garnet theory has been used to determine the effective dielectric function for multi-component systems [23,24]:

$$\varepsilon_{eff} = \varepsilon_{matrix} \frac{1+2\sum_i f_i \frac{\varepsilon_i-\varepsilon_{matrix}}{\varepsilon_i+2\varepsilon_{matrix}}}{1-\sum_i f_i \frac{\varepsilon_i-\varepsilon_{matrix}}{\varepsilon_i+2\varepsilon_{matrix}}} = \varepsilon_{SiO_2} \frac{1+2\left(f_{TiO_2}\frac{\varepsilon_{TiO_2}-\varepsilon_{SiO_2}}{\varepsilon_{TiO_2}+2\varepsilon_{SiO_2}}+f_{Ag}\frac{\varepsilon_{Ag}-\varepsilon_{SiO_2}}{\varepsilon_{Ag}+2\varepsilon_{SiO_2}}\right)}{1-\left(f_{TiO_2}\frac{\varepsilon_{TiO_2}-\varepsilon_{SiO_2}}{\varepsilon_{TiO_2}+2\varepsilon_{SiO_2}}+f_{Ag}\frac{\varepsilon_{Ag}-\varepsilon_{SiO_2}}{\varepsilon_{Ag}+2\varepsilon_{SiO_2}}\right)} \tag{2}$$

where $f_i$ is the filling factor of the *i*th material. $\varepsilon_{eff}$ is a complex function and the complex refractive index dispersion $n = n_{real} + in_{imag}$ has been determined with the aforementioned procedure. In this work we consider that the materials are spatially separated in agreement with the assumptions of the Maxwell-Garnett theory [25].

$\varepsilon_{SiO_2}$ has been determined considering that $\varepsilon_{SiO_2} = n_{SiO_2}^2$ and employing the Sellmeier equation for silicon dioxide taken from Ref. [26]:

$$n_{SiO_2}^2(\lambda) - 1 = \frac{0.6961663\lambda^2}{\lambda^2-0.0684043^2} + \frac{0.4079426\lambda^2}{\lambda^2-0.1162414^2} + \frac{0.8974794\lambda^2}{\lambda^2-9.896161^2} \tag{3}$$

Analogously, $\varepsilon_{TiO_2}$ has been determined considering that $\varepsilon_{TiO_2} = n_{TiO_2}^2$ and employing a refractive index dispersion for titanium dioxide taken from Ref. [27]:

$$n_{TiO_2}(\lambda) = \left(4.99 + \frac{1}{96.6\lambda^{1.1}} + \frac{1}{4.60\lambda^{1.95}}\right)^{1/2} \tag{4}$$

For the dielectric function of Ag nanoparticles the Drude model has been used [28], in which the real and imaginary part of $\varepsilon$ can be written as

$$\varepsilon_1 = \varepsilon_\infty - \frac{\omega_P^2}{(\omega^2+\Gamma^2)} \tag{5}$$

$$\varepsilon_2 = \frac{\omega_P^2 \Gamma}{\omega(\omega^2+\Gamma^2)} \tag{6}$$

For silver the used values of high frequency dielectric function $\varepsilon_\infty$ and carrier damping $\Gamma$ are 9.1 eV and 0.018 eV, respectively [28].

The transfer matrix method has been used to simulate the transmission spectra of the photonic crystals [14,29]. The incident light impinges the sample orthogonally with respect to the surface of the multilayer (with the light coming from the substrate, Figure 1a) and the materials that compose the multilayer are considered paramagnetic. The matrix for the BiVO4 layer is

$$M_{BiVO_4} = \begin{bmatrix} \cos\left(\frac{2\pi}{\lambda}n_{BiVO_4}d_{BiVO_4}\right) & -\frac{i}{n_{BiVO_4}}\sin\left(\frac{2\pi}{\lambda}n_{BiVO_4}d_{BiVO_4}\right) \\ -in_{BiVO_4}\sin\left(\frac{2\pi}{\lambda}n_{BiVO_4}d_{BiVO_4}\right) & \cos\left(\frac{2\pi}{\lambda}n_{BiVO_4}d_{BiVO_4}\right) \end{bmatrix} \tag{7}$$

Where $\lambda$ is the wavelength and $d_{BiVO_4}$ is the thickness of the $BiVO_4$ layers. The matrix of the silica layers doped with titanium dioxide nanoparticles and silver nanoparticles is

$$M_{dSiO_2} = \begin{bmatrix} \cos\left(\frac{2\pi}{\lambda}n_{dSiO_2}d_{dSiO_2}\right) & -\frac{i}{n_{dSiO_2}}\sin\left(\frac{2\pi}{\lambda}n_{dSiO_2}d_{dSiO_2}\right) \\ -in_{dSiO_2}\sin\left(\frac{2\pi}{\lambda}n_{dSiO_2}d_{dSiO_2}\right) & \cos\left(\frac{2\pi}{\lambda}n_{dSiO_2}d_{dSiO_2}\right) \end{bmatrix} \tag{8}$$

Where $d_{SiO_2}$ is the thickness of the doped $SiO_2$ layers and $n_{dSiO_2}$ is the effective refractive index of the doped $SiO_2$ layers determined with Equation 2. The characteristic matrix for the multilayer is

$$M = \begin{bmatrix} m_{11} & m_{12} \\ m_{21} & m_{22} \end{bmatrix} = \prod_{i=1}^{N} M_{BiVO_4} M_{dSiO_2} \quad (9)$$

With $N$ the number of bilayers ($N=10$ in this study). From the matrix $M$ it is possible to determine the transmission coefficient $t$ and the transmission $T$:

$$t = \frac{2n_s}{(m_{11}+m_{12}n_0)n_s+(m_{21}+m_{22}n_0)} \quad (10)$$

$$T = \frac{n_0}{n_s}|t|^2 \quad (11)$$

**Results and Discussion**

The photonic crystals studied are composed by 10 bilayers of BiVO$_4$ and of SiO$_2$ infiltrated with TiO$_2$ nanoparticles and Ag nanoparticles. In this work, the nanoparticle-doped silica layer is called Ag:TiO$_2$:SiO$_2$.

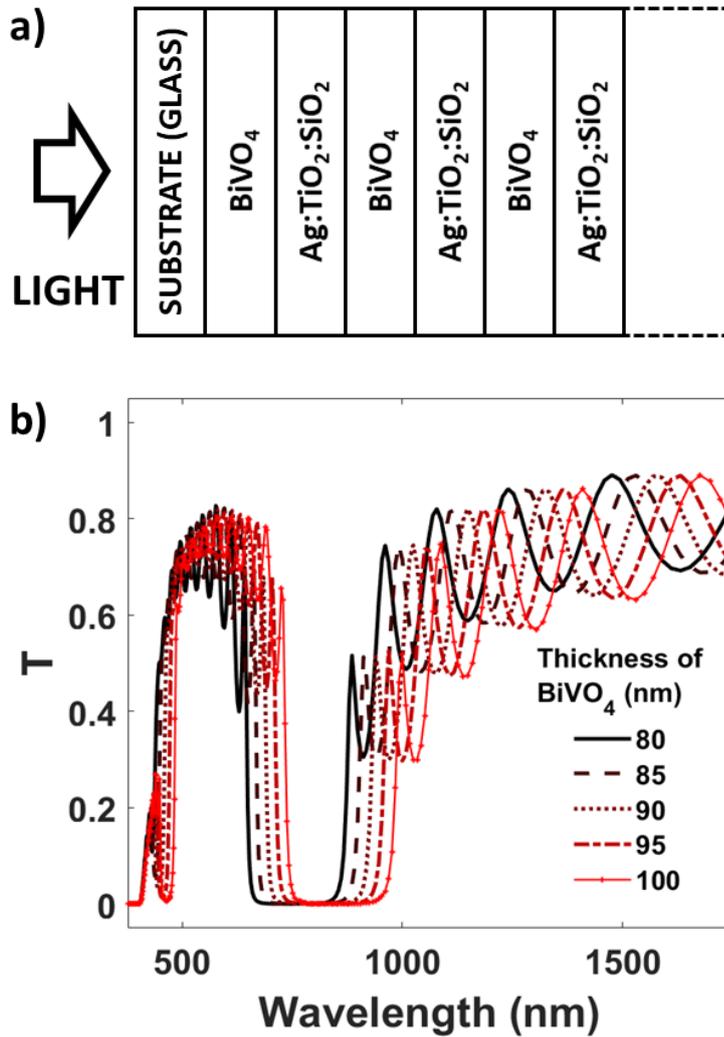

**Figure 1**. a) Sketch of the photonic crystals; b) Transmission spectrum of the BiVO$_4$/Ag:TiO$_2$:SiO$_2$ photonic crystal as a function of the thickness of the BiVO$_4$ layers ($f_{TiO_2,Ag} = 0.05$).

In Figure 1b the transmission spectrum of a BiVO$_4$/Ag:TiO$_2$:SiO$_2$ photonic crystal as a function of the BiVO$_4$ layers thickness is shown. The filling factors for the Ag and TiO$_2$ nanoparticles in the SiO$_2$ layers are 0.05. The thickness of the Ag:TiO$_2$:SiO$_2$ layers is 100 nm. By increasing the BiVO$_4$ layers thickness from 80 nm to 100 nm the red shift of the photonic band gap, initially between 650 nm and 870 nm, is clear. The band gap shift occurs without a remarkable broadening or narrowing in its width.

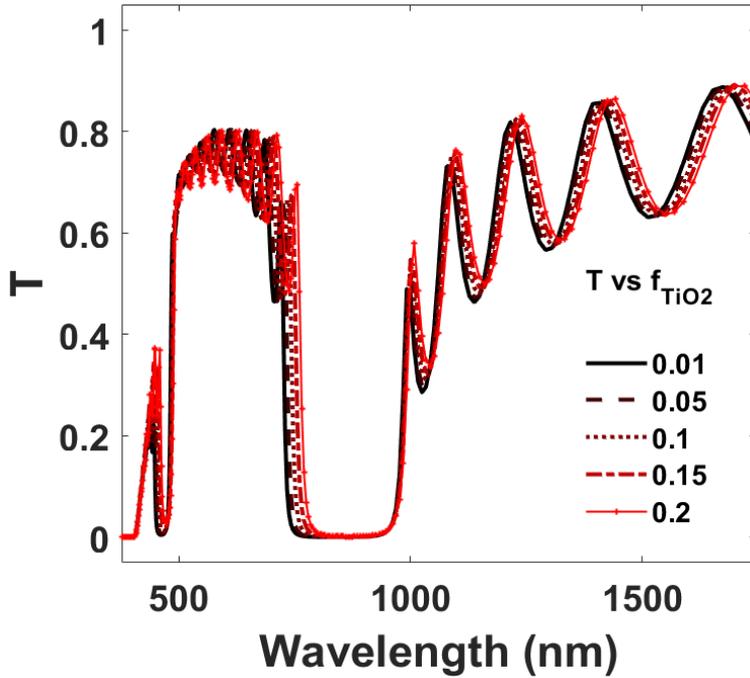

**Figure 2.** Transmission spectrum of the BiVO$_4$/Ag:TiO$_2$:SiO$_2$ photonic crystal as a function of the filling factor of the TiO$_2$ nanoparticles in the Ag:TiO$_2$:SiO$_2$ layers (the thickness of BiVO$_4$ layers is 100 nm and $f_{Ag} = 0.05$).

In Figure 2 the transmission spectrum of the BiVO$_4$/Ag:TiO$_2$:SiO$_2$ photonic crystal as a function of the filing of the TiO$_2$ nanoparticles is depicted. The thickness of BiVO$_4$ layers is 100 nm and $f_{Ag} = 0.05$. With an increase of the filling factor of the titanium nanoparticles from 0.01 to 0.2 a red shift of the band gap, with a concomitant narrowing, is observable.

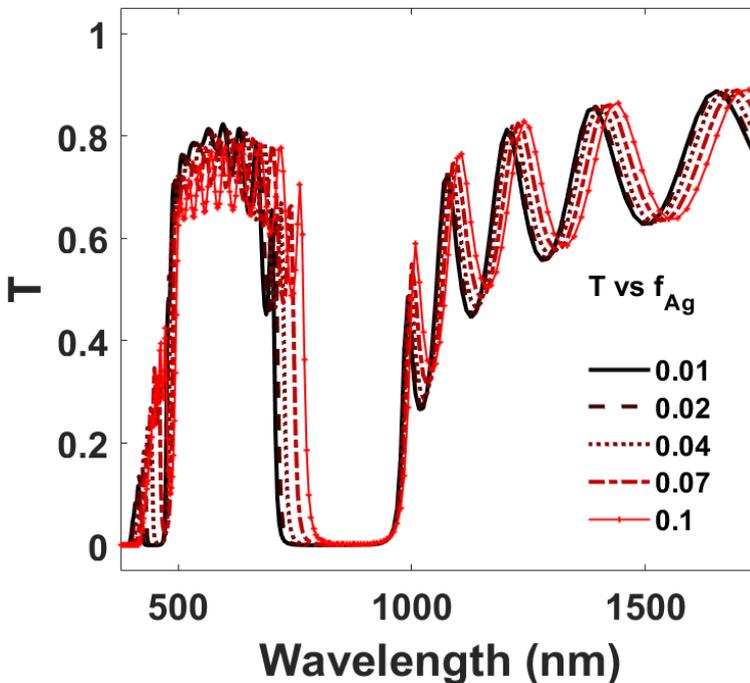

**Figure 3.** Transmission spectrum of the BiVO$_4$/Ag:TiO$_2$:SiO$_2$ photonic crystal as a function of the filling factor of the Ag nanoparticles in the Ag:TiO$_2$:SiO$_2$ layers (the thickness of BiVO$_4$ layers is 100 nm and $f_{TiO_2} = 0.05$).

In Figure 3 the transmission spectrum of the BiVO₄/Ag:TiO₂:SiO₂ photonic crystal as a function of the filing of the Ag nanoparticles is shown. The thickness of BiVO₄ layers is 100 nm and $f_{TiO_2} = 0.05$. Also by increasing the filling factor of silver nanoparticles the red shift of the band gap, concomitant to its narrowing, is significant. With silver nanoparticles, the effect can be obtained with smaller filling factors. Such stronger influence of the silver nanoparticle filling factor in the shift of the photonic band gap edges is highlighted in Figure 4.

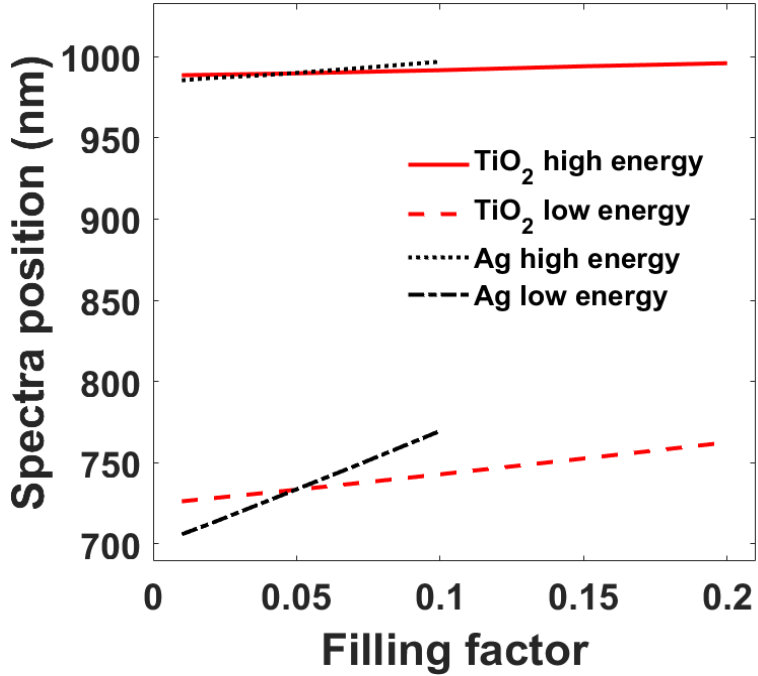

**Figure 4.** Spectral positions of the high and low energy edges of the photonic band gaps as a function of $f_{TiO_2}$ and $f_{Ag}$.

In fact, from Figure 4 it is evident that the concomitant red shift and narrowing results in a negligible shift of the low energy photonic band gap edges (red solid line for titanium dioxide nanoparticle filling factor variation and black dotted line for silver nanoparticle filling factor variation, respectively). With a variation of the filling factor of silver nanoparticles from 0.01 to 0.1 a red shift of 63 nm of the low energy photonic band gap edge has been achieved. In this work the possibility to use many different parameters to manipulate the photonic band gap can open new opportunities to tune, or enhance, the photocurrent and the photocatalytic activity of BiVO₄ [9].

To better understand the possible interaction between the light absorption of BiVO₄ and the optical properties of the photonic crystal it is possible to study the group velocity $v_g$ in the medium. By taking into account the one-dimensional wave equation for periodic Bloch eigenfunctions it is possible to derive the ratio $v_g/c$ [18,30]

$$\frac{v_g}{c} = \frac{\left(\frac{\alpha}{n_2}+\frac{\beta}{n_1}\right)\sqrt{16n_1^2n_2^2-\{(n_1+n_2)^2\cos[2\pi(\alpha+\beta)]-(n_1-n_2)^2\cos[2\pi(\alpha-\beta)]\}^2}}{(n_1+n_2)^2(\alpha+\beta)\sin[2\pi(\alpha+\beta)]-(n_1-n_2)^2(\alpha-\beta)\sin[2\pi(\alpha-\beta)]} \qquad (12)$$

where $\alpha = d_{BiVO_4}n_2/\lambda = d_{BiVO_4}n_{BiVO_4}/\lambda$ and $\beta = d_{dSiO_4}n_2/\lambda = d_{dSiO_4}n_{dSiO_4}/\lambda$. Inside the gap $v_g$ is purely imaginary and is related to the evanescent wave inside the photonic crystal. At the edges of the photonic band gap $v_g$ is real but tends to zero, resulting in an enhancement of the absorption of $BiVO_4$ [31]. With a proper design of the photonic crystal the photocatalytic activity of $BiVO_4$ can be modulated [9].

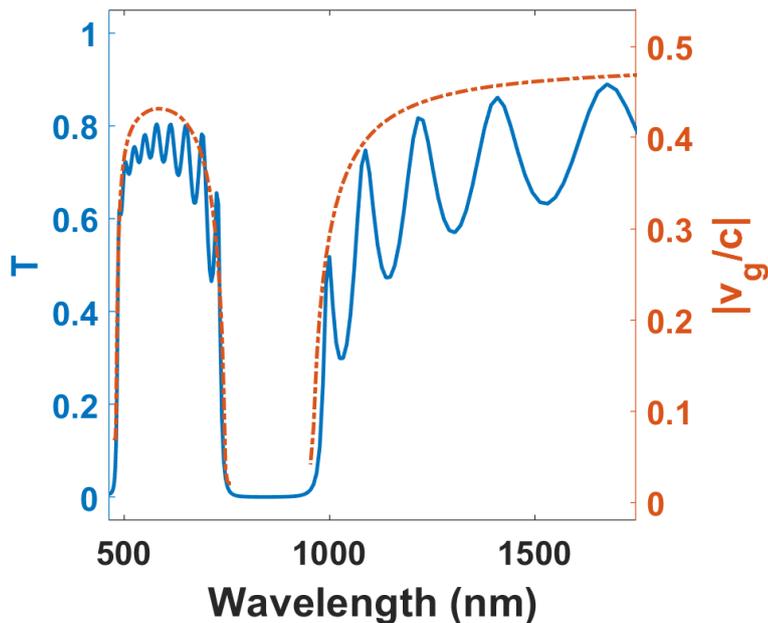

**Figure 5.** Transmission spectrum (blue solid curve) and $v_g/c$ ratio (orange point-dashed curve) of the BiVO$_4$/Ag:TiO$_2$:SiO$_2$ photonic crystal (the thickness of BiVO$_4$ layers is 100 nm, $f_{TiO_2} = 0.05$, and $f_{Ag} = 0.05$).

**Conclusion**
Monoclinic clinobisvanite bismuth vanadate is an interesting material due to its high photocatalytic activity. The inclusion of bismuth vanadate in a photonic crystal gives the possibility to combine the light manipulation and confinement thank to the photonic crystal and the optical properties of BiVO$_4$. In this work, a one-dimensional photonic crystal in which layers of monoclinic clinobisvanite bismuth vanadate and layers of silicon dioxide doped with titanium dioxide nanoparticles and silver nanoparticles has been engineered. The optical properties of the crystal have been studied employing the transfer matrix method, considering the refractive index dispersions of the selected materials. Parameters like the thickness of the bismuth vanadate layers and the filling factor of titanium dioxide nanoparticles and silver nanoparticles significantly influence the spectral position of the photonic band gap. This study can be interesting in order to modulate and filter the absorption of bismuth vanadate layers, for example with a proper overlap between the bismuth vanadate absorption and the photonic band edges where the group velocity of light is slower.

**Acknowledgement**
This project has received funding from the European Research Council (ERC) under the European Union's Horizon 2020 research and innovation programme (grant agreement No. [816313]).